\documentclass[twoside]{article}
\usepackage{amsfonts}
\usepackage{amsmath}
\usepackage{caption}
\usepackage{braket}
\usepackage{multirow}
\usepackage{epsfig}
\usepackage{graphicx}
\usepackage{subfigure}
\usepackage{qic}
\usepackage{calligra}
\usepackage{slashed}
\usepackage[T1]{fontenc}

\begin{document}
\setlength{\textheight}{8.0truein}    

\runninghead{DIRAC FOUR-POTENTIAL TUNINGS-BASED QUANTUM TRANSISTOR} 
            {AGUNG TRISETYARSO}

\normalsize\textlineskip
\thispagestyle{empty}
\setcounter{page}{1}

\copyrightheading{0}{0}{2012}{000--000}

\vspace*{0.88truein}

\alphfootnote

\fpage{1}

\centerline{\bf
DIRAC FOUR-POTENTIAL TUNINGS-BASED QUANTUM TRANSISTOR }
\vspace*{0.035truein}
\centerline{\bf UTILIZING THE LORENTZ FORCE}
\vspace*{0.37truein}
\centerline{\footnotesize
AGUNG TRISETYARSO\footnote{trisetyarso@ittelkom.ac.id}}
\vspace*{0.015truein}
\centerline{\footnotesize\it School of Fundamental Science and Technology, Keio University}
\baselineskip=10pt
\centerline{\footnotesize\it 3-14-1 Hiyoshi, Kohoku-ku, Yokohama-shi, Kanagawa-ken 223-8522, Japan}
\centerline{\footnotesize \it and}
\vspace*{0.015truein}
\centerline{\footnotesize\it Graduate School of Informatics Engineering, Telkom Institute of Technology,}
\baselineskip=10pt
\centerline{\footnotesize\it Jalan Telekomunikasi Terusan Buah Batu, Bandung 40257, Indonesia \footnote{Present affiliation.}}
\vspace*{0.225truein}
\publisher{(received date)}{(revised date)}

\vspace*{0.21truein}

\abstracts{
We propose a mathematical model of \textit{quantum} transistor in which bandgap engineering corresponds to the tuning of Dirac potential in the complex four-vector form. The transistor consists of  $n$-relativistic spin qubits moving in \textit{classical} external electromagnetic fields. It is shown that the tuning of the direction of the external electromagnetic fields generates perturbation on the potential temporally and spatially, determining the type of quantum logic gates. The theory underlying of this scheme is on the proposal of the intertwining operator for Darboux transfomations on one-dimensional Dirac equation amalgamating the \textit{vector-quantum gates duality} of Pauli matrices. Simultaneous transformation of qubit and energy can be accomplished by setting the $\{\textit{control, cyclic}\}$-operators attached on the coupling between one-qubit quantum gate: the chose of \textit{cyclic}-operator swaps the qubit and energy simultaneously, while \textit{control}-operator ensures the energy conservation.
}{}{}

\vspace*{10pt}

\keywords{Dirac Potential, Darboux Transformations, Quantum Transistor, Quantum Logic Gates}
\vspace*{3pt}
\communicate{to be filled by the Editorial}

\vspace*{1pt}\textlineskip    
\section[INTRODUCTION]{\label{sec:level1}INTRODUCTION}

There is an assertion that classical electromagnetic fields can be used to control the quantum behavior\cite{keldysh1965ionization,trisetyarso-2009}. It has the justification from recent experimental works showing that classical fields can be used to excite the population of atom\cite{PhysRevLett.68.2747} or to extremely accelerate the neutral atoms\cite{Eichmann:2009uq}. In this paper, we show that the generation of quantum logic gates is necessary due to the presence of classical electromagnetic fields on a moving \textit{n}-relativistic spin quantum bits. The proposal leads to the possibility to open the definition of \textit{quantum} transistor since the generation of quantum gate is related to the certain bandgap opening.

Bandgap engineering is playing crucial role in classical information theory\cite{tsividis1999operation}. To generate the classical bits in a transistor, the potential should be tuned by step functions determining the conductivity of the gate in transistor: if the potential exceeds threshold of bandgap, the transistor in conductive state meaning generates "\textbf{ON}`` state, while if  it is less than the bandgap threshold, the transistor is in a non-conducting state representing "\textbf{OFF}`` state. Conventionally, these two states are implemented by \textit{constant} functions. Here, we present that the notion of Dirac four-potential can be exploited to emerge the concept of band gap in a quantum transistor: a device used to generate quantum gates due to the presence of electromagnetic fields on the relativistic spin qubit in which the perturbation is represented by Lorentz force on the Dirac Hamiltonian. Contrarily with the classical transistor, the proposed quantum transistor has dependencies of the space-time structures in the Dirac potential. 

Following Moore's Law, the present efforts attempt to minimize the size and increase the capabilities of transistor. However, physical complications, for instance quantum effects such as the Kondo effect  \cite{goldhaber1998kondo} and quantum interference \cite{Anderson:1998uq}, are increasing due to the miniaturization. Several theoretical and experimental frontier efforts have been proposed, in addition to single-electron transistor\cite{kastner1992single}, lateral quantum dot\cite{PhysRevLett.60.848}, single-photon transistor\cite{Chang:2007fk}, and single-molecule optical transistor \cite{Chang:2007fk}, nevertheless the science and technology for a transistor performing quantum computation is still unclear \cite{Ladd:2010kx} .

The strength of quantum computation for solving problem of computation have been known in the recent years \cite{nielsen2002quantum}. Recently, there are several attempts to establish an appropriate theory to encode qubit and to implement quantum computation in realistic physical systems along with measurement-based quantum computing \cite{PhysRevA.68.022312} and adiabatic quantum computing \cite{farhi-2000}. However, theory of how to implement it into the physical system involving the concept of quantum transistor considering \textit{the relativistic effect} is still poorly comprehended. To date, we only find that only Ref. \cite{Hwang:2009sf} is related to quantum transistor in which can potentially implement quantum computation operation for one qubit.

Recent works on quantum computation are constructed based upon Schr\"{o}dinger Hamiltonian at which the potential is scalar. Thus, the perturbations of the potential due to the presence of external forces, such as Zeeman and Stark effects, are expressed in the scalar form. Lately, following the successful application of two-dimensional Dirac equation describing the massless Dirac fermions in graphene\cite{novoselov2005two}, there is a resurgence of interest, so-called supersymmetry quantum mechanics, to solve the problem in quantum mechanics in the one-dimensional Dirac equation as though in ion trapped  \cite{Gerritsma:2010vn} and cavity quantum electrodynamics \cite{trisetyarso-2009}. In this scheme, the Dirac potential is a complex four-vector in which the bases represented in Pauli matrices. Therefore, the perturbation under this scheme is also represented in a vector. Specifically, we are interested to study a Dirac equation possessing its potential in the Lorentz force form as following (for $\hbar=1$)
\begin{equation}
\label{eq00n}
(-i\gamma^{\mu}\partial_{\mu}+\overbrace{q(\vec{E}+\vec{v} \times \vec{B})}^{\text{Dirac potential}})|\psi\rangle=\varepsilon |\psi\rangle.
\end{equation} 
Since the study of quantum computation under the Dirac Hamiltonian is still poorly understood, here we propose the novel technique of the generation of quantum gates by Dirac potential tuning in which it may be used as bandgap engineering for quantum computation.

The change of potential and state in Dirac equation in Eq. (\ref{eq00n}) obeys Darboux transformations implemented by intertwining operations on the Hamiltonian\cite{PhysRevA.43.4602, Nieto2003151}. An important difference from other transforms such as Fourier and Laplace transforms, Darboux transforms creates the new potential and state without changing their domains. Darboux transformations has wide range applications in physics such as to find exact solutions of non-linear Schr\"{o}dinger, sine-Gordon, and Korteweg-de Vries equations\cite{Matveev01}. Darboux transformations for Dirac equations developed in this paper is so-called Bagrov, Baldiotti, Gitman, and Shamshutdinova (BBGS)- Darboux transformations\cite{trisetyarso-2009,bagrov14darboux} in which the Darboux transform has an input of a unitary matrix.
 
\begin{figure}[h]
\centerline{\epsfig{file=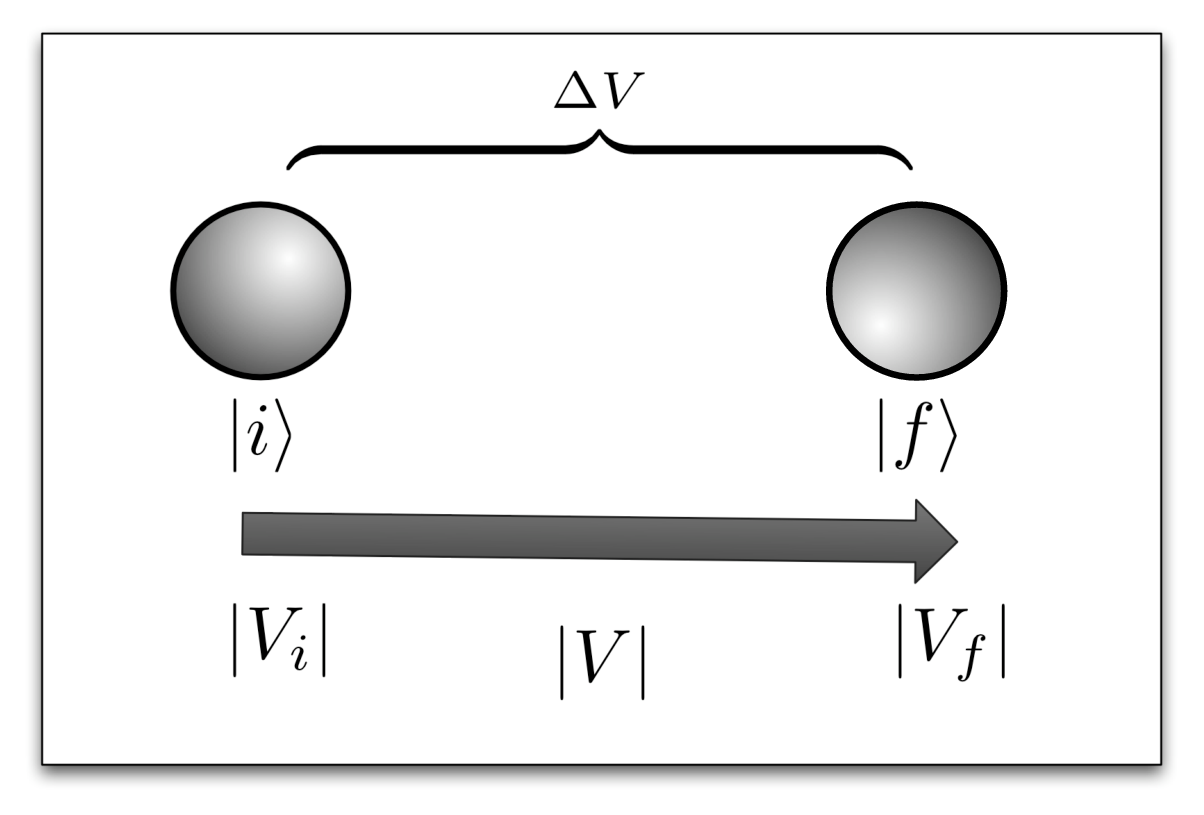, scale=0.40}}
\fcaption{\label{fig:sine}The proposal of voltage bias of one-qubit quantum gate generation. The figure shows the motivation of this paper which is to propose a novel technique of quantum computation transforming the qubits $| initial \rangle \rightarrow |final\rangle$ by tuning the Hamiltonian potential $V \rightarrow V+\Delta V$ in which the $V$ is a complex four-vector representing the bandgap and $\Delta V$ is related to the excitation energy. The transformation is performed by Bagrov, Baldiotti, Gitman, and Shamshutdinova (BBGS)-Darboux transformation\cite{bagrov14darboux} changing $\{V, | initial \rangle\}$ $\rightarrow$ $\{ V+\Delta V, |final\rangle\}$.}
\end{figure}

In this paper, as depicted in Fig. (\ref{fig:sine}), we show how to generate the elementary quantum gates by the tunings of the quantum transistor potential. To do so, the transistor is mathematically expressed by a Dirac Hamiltonian and the tunings are done by varying the operator input, initial potential, and potential differences. In particular, one-dimensional Dirac equation is considered. In graphene, hexagonal stacks of carbon, one-dimensional Dirac equation represents the excitation of massless Dirac-fermions which are subjected to a uniform magnetic field perpendicular to graphene. Here, the plane wave in two-dimensional space is chosen for the solution of (2+1)-dimensional Dirac equation to reduce the problem into one dimension. Mathematical properties underlying this scheme is given in Section (\ref{sec:level1}). The tunings cause perturbation over spatial and temporal regimes of Dirac potential Bloch sphere. Consequently, the generation of unitary gates can be performed by controlling the potential differences. The exposition about this is provided in Section (\ref{sec:level2}).

We outline several new results of this paper: \textit{first}, $n$-relativistic spin qubits moving in the external electromagnetic fields can be exploited for the generation of quantum logic gates. Mathematically, the quantum system is represented by a Dirac equation in which its potential is a Lorentz force in a complex four-vector form and the perturbation obeys Darboux transformation. \textit{Second}, our method suggests a new paradigm of quantum gates: the mathematical expression for perturbing any qubit in the system and the coupling between two gates are well-defined. $\{\sigma_{co}^{\pm}\}$ is the coupling operator for controlling a qubit, $\{\sigma_{cy}^{\pm}\}$ is the operator for cyclely transforming a qubit, and $\alpha \sigma_{0}+(V(t)-\beta(t)) \textbf{U}_{i}$ is the operator for the target qubit. Moreover, the new type of quantum gates are proposed : $\textbf{U-Cyclic}$ gate which can transform the qubits cyclely. 

It is conjectured that the proposal may fit with the novel dual-gate graphene field-effect transistor in which the bandgap is tuned by electrical field\cite{zhang2009direct}.

\section[Mathematical Properties]{\label{sec:level1}Mathematical Properties}

Conventionally, the perturbation theory of a quantum system is described by the approximation methods on Schr\"{o}dinger equation \cite{sakurai1985modern}. In this work, we propose that Darboux transformations on one-dimensional Dirac equations can be used to describe the theory. 

This section consists of three subsections: \textit{first}, Subsection (\ref{sec:level3o}) describes the current problem of so-called the \textit{vector-quantum gate duality} of Pauli matrices. \textit{Second}, the notion of massless Dirac-fermion in graphene is briefly introduced in Subsection (\ref{sec:level3qtp}). \textit{Third}, Subsection (\ref{sec:level3a}) provides the mathematical foundations of Darboux transformations on Dirac equations for multiple qubits. \textit{Fourth}, Subsection  (\ref{sec:level0a}) concerns about the mathematical tools for coupling of two quantum bits.

\subsection[Vector-quantum gates duality of Pauli matrices]{\label{sec:level3o}Vector-quantum gates duality of Pauli matrices}

In this subsection, the \textit{duality} of vector and quantum gate of Pauli matrices is discussed. As widely known, the Pauli matrices, $\{\sigma_{i}|i=0,1,2,3\}$, are the elementary quantum gates in the field of quantum computation\cite{nielsen2002quantum}. Every quantum gates for one qubit operations can be decomposed into the superposition of Pauli matrices as following
\begin{equation}
\label{dcp}
\textbf{U}(g)=\sum_{i=0}^{3}a_{i}\sigma_{i},
\end{equation}
\noindent
where $\textbf{U}(g)$=$\{$$g$$|$ Identity, Hadamard, Pauli-$\{X,Y,Z\}$, Phase, $\frac{\pi}{8}$$\}$ is a unitary matrix, $a_{i}$ is a coefficient of $\sigma_{i}$ in which it has to fulfill the normalization condition $\sum\limits_{i}|a_{i}|^{2}=1$.

On the other side, Pauli matrices can be used as resources for bases of a vector as mentioned in Ref. \cite{0305-4470-22-1-008,PhysRevA.60.785,scott1998complex}. The work in this paper follows the conventions as mentioned in Ref. \cite{scott1998complex} considering the simplicity and the concept is easily understood for wider audiences. The so-called \textit{complex-four vector} is a vector composed by Pauli matrices as the basis vector
\begin{equation}
\label{vct}
\vec{v}=\sum_{i=0}^{3}v_{i}\sigma_{i},
\end{equation}
\noindent
where $v_{i}$ is the coefficient and $\sigma_{i}$ is the basis vector. The basis of complex four-vector, $\{\sigma_{0}, \sigma_{1}, \sigma_{2}, \sigma_{3}\}$, obey the exterior algebra \cite{frankel2004geometry}

\begin{subequations}
\begin{align}
\sigma_{\kappa}~\wedge~\sigma_{\kappa} & = \sigma_{0} \\
\sigma_{\kappa}~\wedge~\sigma_{\gamma}  &= - \sigma_{\gamma}~\wedge~\sigma_{\kappa},
\end{align}
\end{subequations}
\noindent
for $\{\kappa,\gamma\}=1,2,~\text{and}~3$, since the basis can be divided into the scalar part, $\{\sigma_{0}\}$, and the vector part, $\{\sigma_{1}, \sigma_{2}, \sigma_{3}\}$.

The vector conjugate is
\begin{equation}
\label{vct1}
(\vec{v})^{*}=v_{0}\sigma_{0}-\sum_{i=1}^{3}v_{i}\sigma_{i},
\end{equation}
\noindent 
therefore the Lorentz-scalar product easily reads
\begin{equation}
\label{Lvct}
\vec{v}.(\vec{v})^{*}=v^{\mu}g_{\mu \nu}v^{\nu}=(v_{0})^{2}-\sum_{i=1}^{3}(v_{i})^{2},
\end{equation}
\noindent
where $g_{\mu \nu}$ is a Minkowski tensor. The algebraic product of two complex-four vectors is 
\begin{equation}
\label{albpdt}
\vec{a}.(\vec{b})^{*}= a^{\mu}g_{\mu \nu}b^{\nu} \sigma_{0}+((a_{\gamma}b_{0}-b_{\gamma}a_{0})-i\varepsilon_{\alpha \beta \gamma}a_{\alpha}b_{\beta})\sigma_{\gamma},
\end{equation}
\noindent
where $\varepsilon_{\alpha \beta \gamma}$ is Levi-Civita permutation symbol. 

Eq. (\ref{albpdt}) has remarkable features: if the two vectors have same directions, the result is a scalar (or in the direction of $\sigma_{0}$); while, if they have different directions, the result is a complex-four vector. Our model shows that one vector represents the physical system and another one represents the perturbation from environment, thus Eq. (\ref{albpdt}) expresses the interaction between the physical system and environment. The outcomes of the interaction can result a scalar; it may be related to the perturbation due to the presence of Zeeman or Stark effect. Otherwise, it produces a vector, such as found if the perturbation is Lorentz force.

In order to circumvent this discrepancy, we argue that the type of quantum gates in Eq. (\ref{dcp}) is related to the direction of basis vector in Eq. (\ref{vct}). The correspondence can be emerged in the presence of the Lorentz force on the qubits.  The observation of these aspects is provided in the following sections.

\subsection[Massless Dirac-fermions in a uniform magnetic field]{\label{sec:level3qtp}Massless Dirac-fermions in a uniform magnetic field}

Massless Dirac-fermions, massless particles which obey Dirac's (relativistic) equation, can be found on unbound electrons in graphene's honeycomb lattice\cite{novoselov2005two}. The lattice is composed of two inequivalent sublattices labeled \textit{A} and \textit{B}, therefore there are two inequivalent corners \textit{K} and \textit{K'} in the graphene Brillouin zone. These two points are called Dirac points, since the structure of low-energy band at the points corresponds to Dirac cones\cite{RevModPhys.81.109}. 

The wave function of electron which is close to \textit{K}, $\psi(\vec{r})$, is governed by a two dimensional Dirac equation 

\begin{equation}
\label{ep12a}
-iv_{F}\vec{\sigma} \cdot \vec{\nabla}\psi(\vec{r})=\varepsilon \psi(\vec{r}),
\end{equation} 
\noindent
where $v_{F}$ is Fermi velocity\cite{PhysRev.71.622} which has the value $v_{F}\simeq 1 \times 10^{6}~\text{m}~\text{s}^{-1}$ and $\vec{\sigma}=\left(\sigma_{x},\sigma{y}\right)$ are Pauli matrices. The emergence of Dirac cones around Dirac points corresponding to symmetry breaking of energy band in Eq. (\ref{ep12a}), $\varepsilon$, can be explained by representing the Hamiltonian in momentum space 
\begin{equation}
\label{ep112ab}
v_{F}\vec{\sigma} \cdot \vec{k}\psi(\vec{k})=\varepsilon \psi(\vec{k}).
\end{equation} 
If the massless Dirac-fermions in graphene are subjected to a uniform magnetic field, there are degeneracy breaking of energy levels into Landau levels, which are responsible for the half-integer quantum Hall effect \cite{RevModPhys.81.109}. Conventionally, a uniform magnetic field \textit{B} which is perpendicularly applied to the plane of graphene is represented by Landau gauge $\vec{A}$=$B\left(-y,0\right)$. Hence, replacing $-i\vec{\nabla}$ in Eq. (\ref{ep12a}) by $-i\vec{\nabla}+\frac{e\vec{A}}{c}$ and for the wave function in the form $\psi(x,y)=e^{ikx}\phi(y)$, we can obtain
\begin{equation}
\label{ep12ab}
v_{F}\left(\sigma_{y}\partial_{y}-\left(k-\frac{eBy}{c}\right)\sigma_{x} \right)\phi(y)=\varepsilon \phi(y).
\end{equation} 
Eigenvalues corresponding to Landau levels in Eq. (\ref{ep12ab}) can be obtained by changing its form into the same form of atom-field interaction Hamiltonian in Jaynes-Cummings model\cite{jaynes1963comparison}
\begin{equation}
\label{ep12ac}
\left(O\sigma^{+}+O^{\dagger}\sigma^{-}\right)\phi(\xi)=\varepsilon_{1}\phi(\xi),
\end{equation} 
\noindent
where $\text{\calligra{l}}~_{\text{B}}=\sqrt{\frac{c}{eB}}$ is the magnetic length, $\omega_{c}=\sqrt{2}\frac{v_{F}}{\text{\calligra{l}}~_{B}}$ is the cyclotron frequency of the Dirac-fermions, $\varepsilon_{1}=\frac{2\varepsilon}{\omega_{c}}$, $\xi=\frac{y}{\text{\calligra{l}$~_{B}$}}-\text{\calligra{l}}~_{B}k$, 
\noindent
and one-dimensional harmonic-oscillator operators
\begin{subequations}
\begin{equation}
\label{iu9a}
O=\frac{1}{\sqrt{2}}\left(\partial_{\xi}+\xi\right),
\end{equation}
\begin{equation}
\label{iu9b}
O^{\dagger}=\frac{1}{\sqrt{2}}\left(-\partial_{\xi}+\xi\right).
\end{equation}
\end{subequations}

According to Ref.\cite{RevModPhys.81.109}, the $N$-solutions of Eq. (\ref{ep12ab}) can be generated from the zero energy. Therefore, the eigenvectors and their eigenvalues are

\begin{subequations}
\begin{equation}
\label{sop01}
\phi_{N,\pm}(\xi)=\psi_{N-1}\left( \xi \right) \otimes \ket{\uparrow} \pm \psi_{N}\left( \xi \right) \otimes \ket{\downarrow},
\end{equation}
\begin{equation}
\label{iu9b}
\varepsilon_{\pm}\left(N \right)= \pm \omega_{c} \sqrt{N}.
\end{equation}
\end{subequations}
\noindent
where $N=0,1,2,...$ and $\psi_{N}$ is the one-dimensional harmonic oscillator solution. The zero energy state of Landau levels, $N=0$, can only occur if quantum relativistic is taken into account. This particular state is useful to explain the anomalies in the quantum Hall effect.

The above exposition shows how a particular perturbation on massless Dirac-fermions in graphene can explain the anomalies in the quantum Hall effect. In this paper, we introduce an approach in the case of the massless Dirac-fermions are subjected to any directions of an external electromagnetic field to perform quantum computation in graphene.
 
\subsection[Darboux transformations on quantum state and Dirac four-potential]{\label{sec:level3a}Darboux transformations on quantum state and Dirac four-potential}

Darboux transformations for one-dimensional Dirac equation was suggested by Bagrov $et~al.$\cite{bagrov14darboux} and Nieto $et~al.$\cite{Nieto2003151}. Samsonov and Shamshutdinova developed the method to show that it is possible to control the qubit state by the use of an external field\cite{1751-8121-41-24-244023}. Moreover, it was developed in Ref. \cite{trisetyarso-2009} to show the correlation between atomic inversion and Dirac potential in cavity quantum electrodynamics.

Suppose that it is possible to change the potential and the state of a physical system obeying a Dirac equation without changing the energy of the system. Mathematically, the action of  ${\mathcal D}(\textbf{U}_{i})$ on  a set of potential and state of a Hamiltonian is defined by ${\mathcal D}(\textbf{U}_{i})[N]$ $\{V,\Psi\}$ = $\{V[N],\Psi[N]\}$ transforming the old Hamiltonian $\hat{h}(V)\Psi =\varepsilon_{0}\Psi$ $\rightarrow$ $\hat{h}(V[1])\Psi[1] =\varepsilon_{1}\Psi[1]$ $\rightarrow$ ... $\rightarrow$  $\hat{h}(V[N])\Psi[N] =\varepsilon_{N}\Psi[N] $. In the context of quantum physics, it is possible to explain the perturbation phenomena of a quantum system due to the presence of classical effects using Darboux transformations: the action Darboux transformations on a quantum system represented by a set of potential and state, $\{V,\Psi\}$, cause a new set of potential and state of the system, $\{V+\Delta V, \Psi+\Delta \Psi\}$. This transformation is called by Bagrov, Baldiotti, Gitman, and Shamshutdinova (BBGS)- Darboux transformations\cite{trisetyarso-2009,bagrov14darboux}. Here, we modify the transformation by substituting $\sigma_{i}$ by an unitary matrix, $\textbf{U}_{i}$, where for one qubit, $\textbf{U}_{i}$ can represent a vector or $\textbf{U}_{i}\in$ $\{$$\textbf{U}_{0}$= Identity gate, $\textbf{U}_{\{1,2,3\}}$= Pauli-$\{$\textit{X, Y,} and \textit{Z}$\}$ gates, $\textbf{U}_{4}$= Hadamard gate, and $\textbf{U}_{5}$= Phase shift gates$\}$.

We also modify the potential in this paper: it consists of temporal (\textit{scalar}) and spatial (\textit{vector}) terms, thus it is called \textit{Dirac four-potential}. Therefore, it is mathematically defined by $V_{N}(t)=\sum_{j=0}^{3}\sigma_{j}(f_{N}(t))_{j}$. This is a complex four-vector, as defined in Eq. (\ref{vct}), in Minkowskian Bloch sphere and its magnitude is $|V_{N}(t)|=\eta^{jl}(f_{N}(t))_{j}(f_{N}(t))_{l}$, where $\eta^{jl}$ is the metric tensor. 

Because the potential is a complex four-vector, the perturbation also follows to the exterior product or generalized cross product for arbitrary dimensions, instead of dot product. It means that the Dirac potential perturbation has not only magnitude, but also direction, which is quiet different with perturbation in Schr\"{o}dinger potential consisting the magnitude only. The consequences of this choice are shown in the next section.

One can consider to use the \textit{generalized} Pauli-Dirac matrices, {\scriptsize $\overbrace{\theta_{j}}^{n \times n}=\{\overbrace{\sigma_{j}}^{2 \times 2},\overbrace{\gamma_{j}}^{4 \times 4},\overbrace{\pi_{j}}^{8 \times 8},\overbrace{\rho_{j}}^{16 \times 16},...\}$}, where the overbraces denote the dimensions of the matrix, as given in Ref. \cite{Poole:1982mo}, if the space-time dimensions are higher than 4 and also to expand the Dirac equation operating on $n$-qubits. 

The nature of a Dirac equation admits the simulation for two qubits. As shown in Ref. \cite{itzykson1985quantum}, by assuming that the particle is massive\footnote{In other words, the Dirac potential is $-Im$. $I$ is an Identity matrix.}, $m\neq0$, and in the rest frame of particle, $k^{\mu}=(m,\textbf{0})$, the Dirac equation implies that the set of a quantum bit $\{|00\rangle,|01\rangle\}$ is belonged to the positive energy, $\psi^{(+)}$, and the set of quantum state $\{|10\rangle,|11\rangle\}$ corresponds to the negative energy, $\psi^{(-)}$, if the solutions of the Dirac equation are plane wave, $\psi^{(\pm)}=e^{\mp ik^{\mu}x_{\mu}}u_{\pm}(k)$. In this work, we \textit{generalize} the notion into $n$-qubits and also the potential is modified into higher dimensions. Therefore, a \textit{modified} Dirac equation representing from one qubit, $|(\{0,1\}^{1})\rangle$, to $n$-qubits, $|(\{0,1\}^{n})\rangle$, is 
\begin{equation}
\label{gde1}
\begin{array}{cc}
(-i\sigma^{\mu}\partial_{\mu}+\overbrace{\sum\limits_{l=0}^{3}f_{l}\sigma_{l}}^{V})|(\{0,1\}^{1})\rangle&=\varepsilon|(\{0,1\}^{1})\rangle\\ 
\downarrow&\downarrow \\ 
(-i\gamma^{\mu}\partial_{\mu}+ \overbrace{\sum\limits_{l=0}^{3}f_{l}\gamma_{l}}^{V})|(\{0,1\}^{2})\rangle&=\varepsilon|(\{0,1\}^{2})\rangle\\
\downarrow&\downarrow \\ 
(-i\pi^{\mu}\partial_{\mu}+ \overbrace{\sum\limits_{l=0}^{3}f_{l}\pi_{l}}^{V})|(\{0,1\}^{3})\rangle&=\varepsilon|(\{0,1\}^{3})\rangle\\
\downarrow&\downarrow \\ 
(-i\rho^{\mu}\partial_{\mu}+ \overbrace{\sum\limits_{l=0}^{3}f_{l}\rho_{l}}^{V})|(\{0,1\}^{4})\rangle&=\varepsilon|(\{0,1\}^{4})\rangle\\
\downarrow&\downarrow \\ 
... & ... \\
(-i\theta^{\mu}\partial_{\mu}+ \overbrace{\sum\limits_{l=0}^{3}f_{l}\theta_{l}}^{V})|(\{0,1\}^{n})\rangle&=\varepsilon|(\{0,1\}^{n})\rangle\\
\end{array}
\end{equation}

Eq. (\ref{gde1}) has several possibilities of interpretations. According to the original Dirac theory, \textit{n}-qubits would need $\frac{n}{2}$ Dirac-fermions. However, in conformity with the model of massless Dirac-fermions in graphene as shown in Subsection (\ref{sec:level3qtp}), \textit{n}-qubits correspond to \textit{n}-massless Dirac-fermions.

Due to the perturbation, it changes the potential $V_{0}(t)=\sum_{j}\sigma_{j}(f_{0}(t))_{j} \rightarrow V_{1}(t)=\sum_{j}\sigma_{j}(f_{1}(t))_{j} \rightarrow ... \rightarrow V_{N}(t)=\sum_{j}\sigma_{j}(f_{N}(t))_{j}$ obeying the intertwining operation 
\begin{equation}
\label{ch0q}
\hat{{\mathcal L}}(\textbf{U}_{i})\hat{h}(V_{0})=\hat{h}(V_{1})\hat{{\mathcal L}}(\textbf{U}_{i})\rightarrow \hat{{\mathcal L}}(\textbf{U}_{i})\hat{h}(V_{1})=\hat{h}(V_{2})\hat{{\mathcal L}}(\textbf{U}_{i}) \rightarrow ... \rightarrow \hat{{\mathcal L}}(\textbf{U}_{i})\hat{h}(V_{N-1})=\hat{h}(V_{N})\hat{{\mathcal L}}(\textbf{U}_{i}).
\end{equation}
\noindent
The chain of relations in Eq. (\ref{ch0q}) means that the new potential, from $V_{1}$ to $V_{N}$, can be generated under intertwining operations from the initial potential ($V_{0}$). The intertwining operator is 
\begin{equation}
\hat{\mathcal L}(\textbf{U}_{i})=\frac{d}{dt}+\hat{\mathcal B}(\textbf{U}_{i}),
\end{equation}
\noindent
where

\begin{equation}
\label{e0}
\hat{\mathcal B}(\textbf{U}_{i}) = \alpha_{i}(t)\sigma_{0}+(V_{N-1}(t)-\beta_{i}(t))\textbf{U}_{i}.
\end{equation} 
\noindent
In Eq. (\ref{e0}), $\textbf{U}_{i}$ is a vector represented in matrix form. These also lead the \textit{linear} transformation of state\cite{bagrov14darboux}, for one qubit, $\Psi \rightarrow \Psi[1]=\hat{\mathcal L}(\textbf{U}_{i}) \Psi \rightarrow \Psi[1]=\hat{\mathcal L}(\textbf{U}_{i}) \Psi[1] \rightarrow ... \rightarrow \Psi[N]=\hat{\mathcal L}(\textbf{U}_{i}) \Psi[N-1]$.

\noindent

It is clear that the terms $\{\alpha_{i}, \beta_{i}\}$ represent the perturbation terms. To conclude about this method, in other words, the change of Dirac \textit{potential} under BBGS-Darboux transformation obeys \textit{intertwining operations}, while the change of \textit{state} follows \textit{linear transformations} \cite{Nieto2003151}.

\subsection[Control operators]{\label{sec:level0a}Control and cyclic operators}

In this subsection, we introduce the two operators which will be extensively used in constructing quantum circuit. These operators satisfy idempotent operation, ${\mathcal A}^{2}={\mathcal A}$. $\textit{First}$, $\textit{control operators}$, $\{\sigma^{(\pm)}_{(0,3)}=\sigma^{(\pm)}_{(co)}=\frac{\sigma_{0}\pm \sigma_{3}}{2}\}$. $\textit{Second}$, $\textit{cyclic operators}$, $\{\sigma^{(\pm)}_{(1,2)}=\sigma^{(\pm)}_{(cy)}=\frac{\sigma_{1}\pm i\sigma_{2}}{2}\}$, which are widely known as matrix representations of the raising and lowering operators \cite{sakurai1985modern}. Let us define:
\begin{subequations}
\label{new1a}
\begin{align}
{\mathcal A}\bigoplus_{co\{\pm\}} {\mathcal B}=\sigma^{(+)}_{(co)}\otimes{\mathcal A}+\sigma^{(-)}_{(co)}\otimes{\mathcal B},
\end{align}
\begin{align}
{\mathcal A}\bigoplus_{cy\{\pm\}} {\mathcal B}={\mathcal A}\otimes\sigma^{(+)}_{(cy)}+{\mathcal B}\otimes\sigma^{(-)}_{(cy)}.
\end{align}
\end{subequations}
It is shown below that these formalisms are useful to construct any unitary matrices and to reduce a lot of space to write multiple qubit operations. From these operators, one can obtain the following gates for two qubits:

\begin{enumerate}
\item $\textbf{SWAP}$-gate = $\sigma^{(+)}_{(co)}\otimes \sigma^{(+)}_{(co)} + \sigma^{(-)}_{(co)}\otimes \sigma^{(-)}_{(co)}+\sigma^{(+)}_{(cy)}\otimes \sigma^{(-)}_{(cy)} + \sigma^{(-)}_{(cy)}\otimes \sigma^{(+)}_{(cy)}$
\\$=\sigma^{(+)}_{(co)}\bigoplus\limits_{co\{\pm\}} \sigma^{(-)}_{(co)}+\sigma^{(-)}_{(cy)}\bigoplus\limits_{cy\{\pm\}} \sigma^{(+)}_{(cy)}$
\\$=\left( \begin{matrix} 1&0&0&0\\ 0&0&1&0\\0&1&0&0\\0&0&0&1 \end{matrix} \right).$

\item $\textbf{Full~SWAP}$-gate $= \sigma^{(+)}_{(cy)}\otimes \sigma^{(+)}_{(cy)} + \sigma^{(-)}_{(cy)}\otimes \sigma^{(-)}_{(cy)}+\sigma^{(+)}_{(cy)}\otimes \sigma^{(-)}_{(co)} + \sigma^{(-)}_{(cy)}\otimes \sigma^{(+)}_{(co)}$
\\$=\sigma^{(-)}_{(co)}\bigoplus\limits_{cy\{\pm\}} \sigma^{(+)}_{(co)}+\sigma^{(+)}_{(cy)}\bigoplus\limits_{cy\{\pm\}} \sigma^{(-)}_{(cy)}$
\\$=\left( \begin{matrix} 0&0&0&1\\ 0&0&1&0\\0&1&0&0\\1&0&0&0 \end{matrix} \right).$
\\
\noindent
It is named by $\textbf{Full~SWAP}$-gate, because it transforms $\ket{00}\leftrightarrow \ket{11}$ and $\ket{01}\leftrightarrow \ket{10}$.
\end{enumerate}

Let us consider the use of $\textit{control}$- and  $\textit{cyclic}$-operators for constructing the quantum gates.

\begin{enumerate}
\item \begin{eqnarray}\label{eq0a}\textbf{Controlled-U}~\text{gate} = \sigma^{(+)}_{(co)}\otimes \sigma_{0} + \sigma^{(-)}_{(co)}\otimes \textbf{U}_{i} \nonumber
\\ = \sigma_{0}\bigoplus_{co\{\pm\}} \textbf{U}_{i}.~~~~~~~~~~~~~~
\end{eqnarray}
$~~~~~~~~~~~~~~~~~~~~~~~~=\left( \begin{array}{c|c}\begin{array}{cc}1 & 0 \\ 0 & 1\end{array} & \begin{array}{cc}0 & 0 \\ 0 & 0\end{array} \\ \hline  \begin{array}{cc}0 & 0 \\ 0 & 0\end{array} & \textbf{U}_{i}\end{array} \right)$.
\item \begin{eqnarray}\label{q1}\textbf{U-Cyclic}~\text{gate} = \sigma_{0} \otimes \sigma^{(\pm)}_{(cy)}+ \textbf{U}_{i} \otimes \sigma^{(\mp)}_{(cy)} \nonumber
\\= \sigma_{0}\bigoplus_{cy\{\pm,\mp\}} \textbf{U}_{i}.~~~~~~~~~~~~
\end{eqnarray}\\
\indent

If $\textbf{U}_{i}$ is $\sigma_{1}$ and by choosing Eq. (\ref{q1}) in the form $\sigma_{0} \otimes \sigma^{(-)}_{(cy)}+ \textbf{U}_{i} \otimes \sigma^{(+)}_{(cy)}= \sigma_{0}\bigoplus_{cy\{\mp\}} \textbf{U}_{i}$, the gate is called $\textit{clockwise-cyclic}$-gate (\textbf{CC}-gate),  because it has the form
\begin{equation}\left( \begin{array}{c|c}\begin{array}{cc}0 & 0 \\ 1 & 0\end{array} & \begin{array}{cc}0& 1 \\ 0 & 0\end{array} \\ \hline \begin{array}{cc}0 & 1 \\ 0& 0\end{array} & \begin{array}{cc}0 & 0 \\ 1 & 0\end{array} \end{array} \right)\end{equation}

and it transforms 

\begin{equation}
\begin{array}{l}
\displaystyle \ket{00} \rightarrow \ket{01}\\
\displaystyle  \ket{01} \rightarrow \ket{10}  \\
\displaystyle \ket{10}  \rightarrow \ket{11}  \\
\displaystyle  \ket{11}  \rightarrow \ket{00}. 
\end{array} 
\label{xdef}
\end{equation}

The $\textit{counterclockwise-cyclic}$-gate (\textbf{CCC}-gate) can be performed by the substitution of $\textbf{U}_{i}$ by $\sigma_{1}$ and Eq. (\ref{q1}) in the form $\sigma_{0} \otimes \sigma^{(+)}_{(cy)}+ \textbf{U}_{i} \otimes \sigma^{(-)}_{(cy)}= \sigma_{0}\bigoplus\limits_{cy\{\pm\}} \textbf{U}_{i}$. It has the form as following
\begin{equation}\left( \begin{array}{c|c}\begin{array}{cc}0 & 1 \\ 0 & 0\end{array} & \begin{array}{cc}0& 0 \\ 1 & 0\end{array} \\ \hline \begin{array}{cc}0 & 0 \\ 1& 0\end{array} & \begin{array}{cc}0 & 1 \\ 0 & 0\end{array} \end{array} \right)\end{equation}

and it changes

\begin{equation}
\begin{array}{l}
\displaystyle \ket{00} \rightarrow \ket{11}\\
\displaystyle  \ket{11} \rightarrow \ket{10}  \\
\displaystyle \ket{10}  \rightarrow \ket{01}  \\
\displaystyle  \ket{01}  \rightarrow \ket{00}. 
\end{array} 
\label{xdef2}
\end{equation}

\end{enumerate}

\begin{figure}[h!]
\begin{center}
\subfigure[~Illustration for $\textit{clockwise-cyclic}$ operations of two qubits.]{\epsfig{file=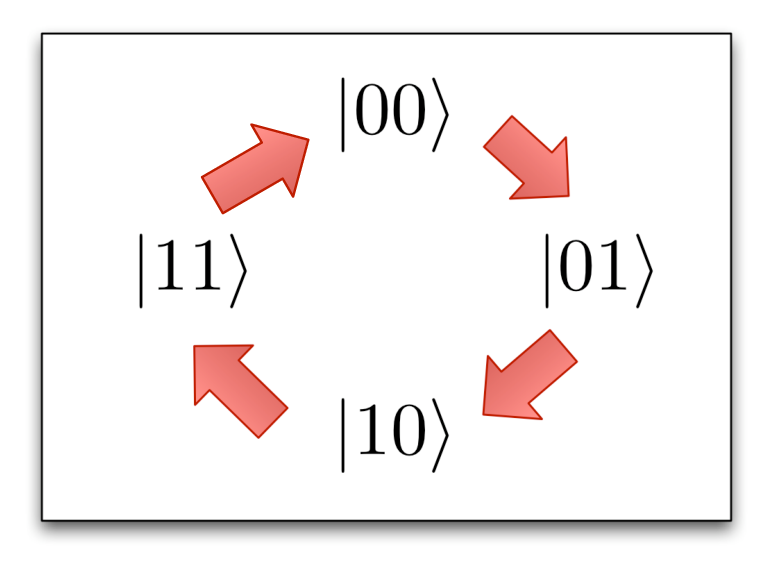, scale=0.45}}
\subfigure[~Illustration for $\textit{counterclockwise-cyclic}$ operations of two qubits.]{\epsfig{file=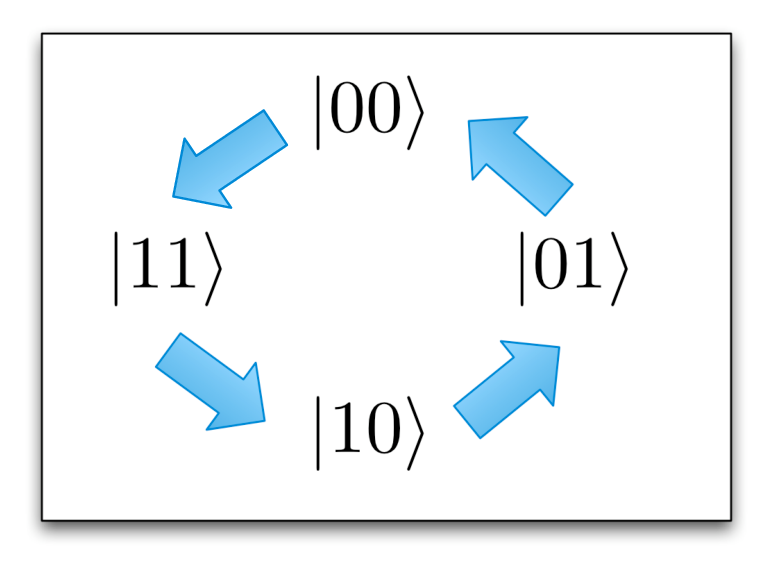, scale=0.45}}
\fcaption{\label{cc1} The illustration of how the use of $\textbf{U-Cyclic}$ for quantum computation is. If $\textbf{U}_{i}$ is $\sigma_{1}$, one can obtain $\textit{clockwise-cyclic}$ and $\textit{counterclockwise-cyclic}$ operations of two qubits. }
\end{center}
\end{figure}
\noindent
In other words, $\left(\textbf{C}\textbf{C}\right)^{4}=\left(\textbf{C}\textbf{C}\textbf{C}\right)^{4}=1$ is to perform one round rotation as shown in Fig. (\ref{cc1}). Interestingly, one can also find the following relations
\begin{subequations}\label{wrt0}
\begin{equation}\label{wrt1a} \left(\textbf{C}\textbf{C}\right)=\left(\textbf{C}\textbf{C}\textbf{C}\right)^{\dagger}=\left(\textbf{C}\textbf{C}\textbf{C}\right)^{-1}, \end{equation}
\begin{equation}\label{wrt1b}  \left(\textbf{C}\textbf{C}\textbf{C}\right)=\left(\textbf{C}\textbf{C}\right)^{\dagger}=\left(\textbf{C}\textbf{C}\right)^{-1}, \end{equation}
\end{subequations}  
\noindent
which are the properties of unitary matrices, besides the facts that |det(\textbf{C}\textbf{C})|=|det(\textbf{C}\textbf{C}\textbf{C})|=1.

The definitions in Eq. (\ref{new1a}) can also be used to represent the standard sets of universal gates of one qubits
\begin{subequations}
\label{un1}
\begin{equation}
\textbf{Hadamard}~\text{gate}=\frac{1}{\sqrt{2}} \left(\sigma_{(co)}^{(+)}-\sigma_{(co)}^{(-)}+\sigma_{(cy)}^{(+)}+\sigma_{(cy)}^{(-)} \right),
\end{equation}
\begin{equation}
\textbf{Phase}~\text{gate}=\sigma_{(co)}^{(+)}+i\sigma_{(co)}^{(-)},
\end{equation}
\begin{equation}
\frac{\textbf{$\pi$}}{4}~\text{gate}=\sigma_{(co)}^{(+)}+e^{i\frac{\pi}{4}}\sigma_{(co)}^{(-)},
\end{equation}
\end{subequations}
\noindent
and for two qubits
\begin{equation}
\label{cnot1a}
\textbf{Controlled-NOT}~\text{gate}=\overbrace{\sigma_{0}}^{\text{Controlled}} \bigoplus\limits_{co(\pm)} \overbrace{\sigma_{1}}^{\text{NOT}}.
\end{equation}
\noindent
The application of the approach in this Subsection to the rest of discussion is given in Subsection (\ref{sec:level2tu}).

\section[Quantum gates generation by a quantum transistor]{\label{sec:level2}Quantum gates generation by a quantum transistor}

The relativistic electron theory of Dirac successfully explains the quantum phenomena such as the $g$ factor and the proper fine structure of Zeeman effect correctly \cite{Jackson:1999fk}. Several efforts have been achieved for a better understanding of Dirac theory by modifying the equation. The famous one was presented by Foldy and Wouthuysen which was called by Foldy-Wouthuysen transformation afterwards \cite{PhysRev.78.29}; they showed a kind of canonical transformation in which can connect between Dirac (or relativistic quantum) and Pauli (non-relativistic quantum) theory and to generate \textit{Zitterbewegung}. Nevertheless, it is still elusive what kind of transformation on Dirac equation can be used to implement quantum computation on Dirac theory. Motivated by supersymmetric quantum mechanics, we shall present the technique to exploit the one-dimensional Dirac equation into the field of quantum computation. 

Suppose that it is necessary to transform a Dirac equation possessing a classical perturbation as following

\begin{equation}
\label{1Dqa}
\overbrace{(-i\theta^{\mu}\partial_{\mu}+V)}^{\hat{h}_{0}(V)}|\{0,1\}^{n}\rangle_{i}=\varepsilon_{0}|\{0,1\}^{n}\rangle_{i} \rightarrow \overbrace{(-i\theta^{\mu}\partial_{\mu}+V+\Delta V)}^{\hat{h}_{1}(V+\Delta V)}|\{0,1\}^{n}\rangle_{f}=\varepsilon_{1}|\{0,1\}^{n}\rangle_{f}
\end{equation} 
\noindent
where $\{V,|\{0,1\}^{n}\rangle_{i}\}$ is a  \textit{initial} set of potential and qubit and  $\{V+\Delta V,|\{0,1\}^{n}\rangle_{f}\}$ is a \textit{final} set of  potential and qubit. The matrix $\theta^{\mu}$ is a generalized form of Pauli-Dirac matrix as explained in Ref. \cite{Poole:1982mo}, $\theta^{\mu}$ = $\{\sigma^{\mu},\gamma^{\mu},\rho^{\mu},...\}$. Therefore, Eq. (\ref{1Dqa}) provides a correlation between Dirac potential and $n$-qubits quantum computation if one-fold Darboux transformation on the equation is necessary to be defined.

Below, we present that this kind of transformation can be realized by the action of BBGS-Darboux transformation as introduced in Subsection (\ref{sec:level3a}) on the equation possessing the complex four-vector potential. It is shown below that quantum gates can be generated due to the perturbation of the potential of $n$-relativistic spin qubit moving on classical electromagnetic field. The exposition for a single qubit, two qubits, and followed by a generalization to multi-qubit systems are given; also, another way to construct a quantum circuit, i.e., in the case of coupling one relativistic qubit, is presented.

\subsection[One qubit quantum gates generation by a quantum transistor]{\label{sec:level2a0}One qubit quantum gates generation by a quantum transistor}

In this scheme, initially the particle is at rest and has the Dirac potential $V_{0}$ and in quantum state $|\{0,1\}^{1}\rangle_{i}$. This set of Dirac potential and state, $\{V_{0},|\{0,1\}^{1}\rangle_{i}\}$, belongs to the following one-dimensional stationary Dirac equation
\begin{equation}
\label{eq1}
\hat{h}_{0}|\{0,1\}^{1}\rangle_{i} =\varepsilon_{0}|\{0,1\}^{1}\rangle_{i},
\end{equation}
where $\hat{h}_{0}$ = ($i\sigma_{z}\frac{d}{dt}$+$V_{0}(t)$), $V_{0}(t)=\sum_{j}\sigma_{j}(f_{0}(t))_{j}$, and $\varepsilon_{0}$ is a constant. 

The unitary transformation on the initial quantum state resulting the new quantum state is achieved if the relativistic qubit moves in the external electromagnetic fields so that the system has a set of new potential and state, $\{V_{0}+\Delta V, |\{0,1\}^{1}\rangle_{f}\}$, which belongs to the new Dirac Hamiltonian
\begin{equation}
\label{eq1a}
\hat{h}_{1}|\{0,1\}^{1}\rangle_{f}=\varepsilon_{1}|\{0,1\}^{1}\rangle_{f},
\end{equation}
where $\hat{h}_{1}$ = ($i\sigma_{z}\frac{d}{dt}$+$V_{f}(t)$), $V_{1}(t)$=$V_{0}+\Delta V$=$\sum_{j}\sigma_{j}(f_{1}(t))_{j}$, and $\varepsilon_{1}$ is a constant. 

Let us assume that the potential perturbation, $\Delta V$, is related to a relativistic qubit with charge in external electromagnetic fields, hence the new potential is in the following form
\begin{subequations}
\label{rlH}
\begin{align}
V_{1}&=V_{0}+\Delta V \nonumber
\\&=V_{0}+\frac{\dot{\vec{p}}}{|\dot{\vec{p}|}},
\end{align}
\end{subequations}
\noindent 
where  $\dot{\vec{p}}=\sum\limits_{\alpha=0}^{3}(\dot{p})^{\alpha}\sigma_{\alpha}$ reads\cite{Jackson:1999fk}
\begin{subequations}
\begin{align}
\label{lrntz}
(\dot{p})^{\alpha}&=qv_{\beta}F^{\alpha \beta}\nonumber
\\&=qv_{\beta}(\partial^{\alpha}A^{\beta}-\partial^{\beta}A^{\alpha}) 
\end{align}
\end{subequations}
\noindent
and where $q$= charge of the qubit, $v_{\beta}=(v_{0},-v_{1},-v_{2},-v_{3})$ is the four-velocity of the qubit,  $A^{\beta}=(A^{0},A^{1},A^{2},A^{3})$ is the four-potential of the external electromagnetic fields, and $F^{\alpha \beta}$ is the antisymmetric field-strength tensor. In the complex-four vector representation, the Lorentz force is in the following form\cite{scott1998complex,PhysRevA.60.785}

\begin{subequations}
\label{lrntzcv}
\begin{align}
\dot{\vec{p}}&= -q\bigg(\frac{\vec{v}.\square \vec{A}+\overline{\square \vec{A}}.\vec{v}}{2}\bigg) \nonumber \\
&=\text{Re}(\vec{v}.\square \vec{A})
\end{align}
\end{subequations}
\noindent

Therefore, the representation of Eq. (\ref{eq00n}) in a complex four-vector form is

\begin{equation}
\label{neq00}
(-i\theta^{\mu}\partial_{\mu}+\sum\limits_{k=0}^{3}f_{k}\theta_{k})|\{0,1\}^{n}\rangle_{i}=\varepsilon_{0}|\{0,1\}^{n}\rangle_{i}.
\end{equation}

For the case of a single qubit, the perturbation is represented by the exterior product between two complex four-vector represented by $2 \times 2$ matrix as following, \begin{equation}\label{pert1}\Delta V=-i\vec{\sigma_{z}} \wedge \vec{\textbf{U}_{i}}, \end{equation}  
\noindent
where the $\vec{\sigma_{z}}$ belongs to the direction of relativistic spin qubit and $\vec{\textbf{U}_{i}}$ is related to the direction of external electromagnetic fields. Therefore, the direction of external classical electromagnetic fields is an input of the perturbing potential.

The proposed scheme covers potential initialization and switching. In the context of dual-gate graphene field-effect transistor, the initial Dirac potential may be related to the perpendicular electric field generating the bandgap\cite{zhang2009direct} and the switching Dirac potential corresponds to the energy transition of the electrons needed for the excitation from the valence band to the conduction band. It is shown in the theorem below that the process of potential initialization and switching is unique for every quantum gate generation. Following exposes the detailed explanation of the scheme.
\\
\noindent
\\
$\textbf{Theorem~1}$. \textit{Suppose the initial potential is $V_{0}(t)=\sum_{i}\sigma_{i}(f_{0}(t))_{i}$. The one fold BBGS-Darboux transformations on the equation (\ref{eq1}) at which the final potential is $V_{1}(t)=V_{0}(t)+\Delta V, where~\Delta V= -i\vec{\sigma_{z}}\wedge\vec{\textbf{U}_{i}}$,  is a suffice condition for} $\hat{\mathcal L}(\{\textbf{U}_{i}\})=\{\textbf{U}_{i}\}.$ $(V_{0}(t))_{0}$ \textit{and} $(V_{1}(t))_{0}$ \textit{are the vector variables.}

\textit{Proof}. Consider 
\begin{equation}
\label{eq2}
\hat{\mathcal L}(\textbf{U}_{i})\hat{h}_{0}\Psi=\hat{h}_{1}\hat{\mathcal L}(\textbf{U}_{i})\Psi.
\end{equation}

One can find

\begin{subequations}
\label{eq7s}
\begin{align}
\label{eq7sa}
\sigma_{0}(V_{0}(t)-V_{0}(t))+i[\textbf{U}_{i},\sigma_{z}](V_{0}(t)-\beta_{i}(t)-1)=0~
\end{align}
\begin{align}
\label{eq7sb}
i\sigma_{z}\textbf{U}_{i}(\dot{\beta}_{i}(t)-\dot{V}_{0}(t)-\alpha_{i})+\alpha_{i}\sigma_{0}(V_{0}(t)-V_{0}(t))+ (V_{0}(t)-\beta_{0}(t))(V_{0}(t)-(V_{0}(t)\textbf{U}_{i}~~~~~~~~~~~~~~~~~~~~~~~~~~~~~~~~~~~~~~~~~~~~~~~\nonumber
\\+i\sigma_{z}((\beta_{0}(t)-V_{0}(t)-\dot{\alpha}_{i}(t))+\sigma_{0}\dot{V}_{0}(t).~~~~~~~~~~~~~~~~~~~~~~~~~~~~~~~~~~~~~~~~~~~~~~~~~~~~~~~~~~~~~~~~~~~~~~~~~~
\end{align}
\end{subequations}
\noindent

Let us examine the constraints of how to generate $\textbf{U}_{i}$ gate. By setting $\alpha_{i}$ up to 1, the resume for $\textbf{U}_{i}$ can be obtained as shown in Tabs. (\ref{tab:1w}), (\ref{tab:1e}), (\ref{tab:1r}), and (\ref{tab:1t}), respectively.

The illustrations show the initial potential preparations (\textit{blue} dashed line) followed by the tuning of final potential (\textit{red} dashed line) to generate a certain quantum gates. The tunings can be accomplished by several manners: \textit{first}, temporal displacement of the initial potential, \textit{second}, displacement on the potential magnitude, and, \textit{third}, by changing the form of the initial potential function. Under the scheme, some light is cast on the question of how a quantum transistor achieves quantum computation by an examination of the correlation between Dirac potentials and quantum gate generation. 

\begin{figure}[h]
\centerline{\epsfig{file=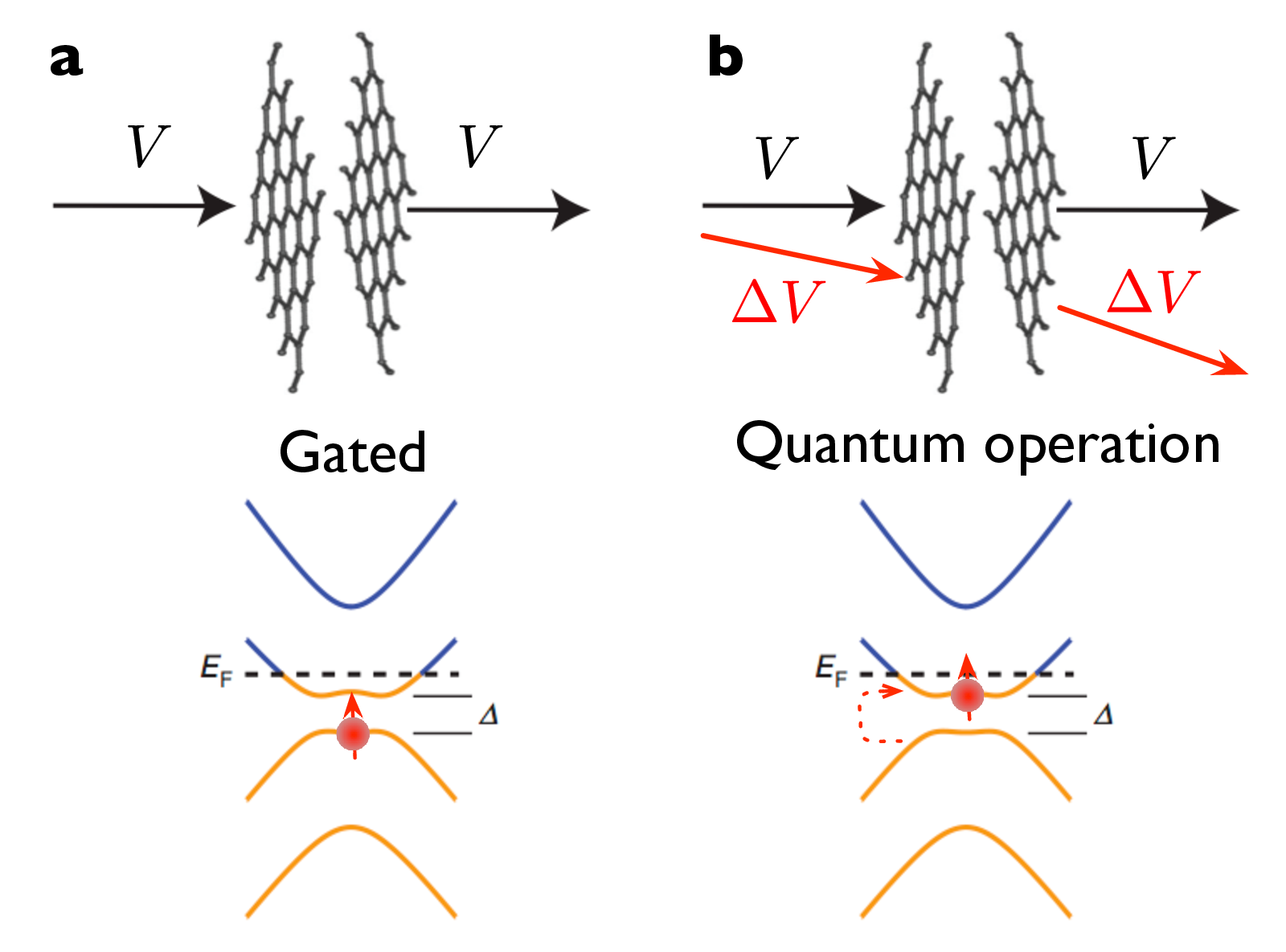, scale=0.40}}
\fcaption{\label{gfet} The illustration of the possibility for the implementation in novel dual-gate graphene FETs\cite{zhang2009direct}: \textbf{a}, A non-zero bandgap $\Delta$ and a the Fermi energy shift, $E_{F}$, are induced by the electric fields. \textbf{b}, The perturbation on Dirac potential, $\Delta V$, excites the electron from the valence band to the conduction band and performs the quantum computation.}
\end{figure}

As shown in Tab. (\ref{tab:1w}), the \textbf{IDENTITY}-gate can be realized by preparing the initial potential in the magnitude of parabolic form. The structure of potential function is transformed from time-dependency to space-time dependencies due to the magnitude displacement of the potential. It is similar to the case of quantum state transition due to carrier-photon scattering event in intraband of semiconductor\cite{harrison2005quantum}: the relativistic spin qubit absorbs the light during the moving. By assuming the photon momentum to be zero, the transition on the band diagrams are always vertical. As illustrated in Fig. (\ref{gfet}), in the case of dual-gate graphene field-effect transistor, the parabolic Dirac potential yields the bandgap, then the electron is excited to the conduction band due to the tuning of electromagnetic field. 

The \textbf{NOT}-gate in $z$-basis can be performed by temporal displacement of the initial to the final potential in the parabolic form. The resume of this scheme is given in Tab. (\ref{tab:1e}).

Tab. (\ref{tab:1r}) provides the scheme to generate $\sigma_{y}$-Pauli matrix. It can be realized by sharpening the initial parabolic potential followed by temporal and magnitude displacements of the potential.

Furthermore, the $\sigma_{z}$-Pauli matrix which is \textbf{NOT}-gate of phase state can be accomplished by setting the initial potential up in constant function, followed by tuning the final potential in parabolic form.   

The other type of quantum gates can be determined by similar manners.

\indent
\begin{center}
\begin{table}[h]
\begin{center}
\begin{tabular}{|c|c|c|c|}
\multicolumn{4}{c}{ $\textbf{Potential dynamics}$} \\ 
\multicolumn{4}{c}{\centerline{\epsfig{file=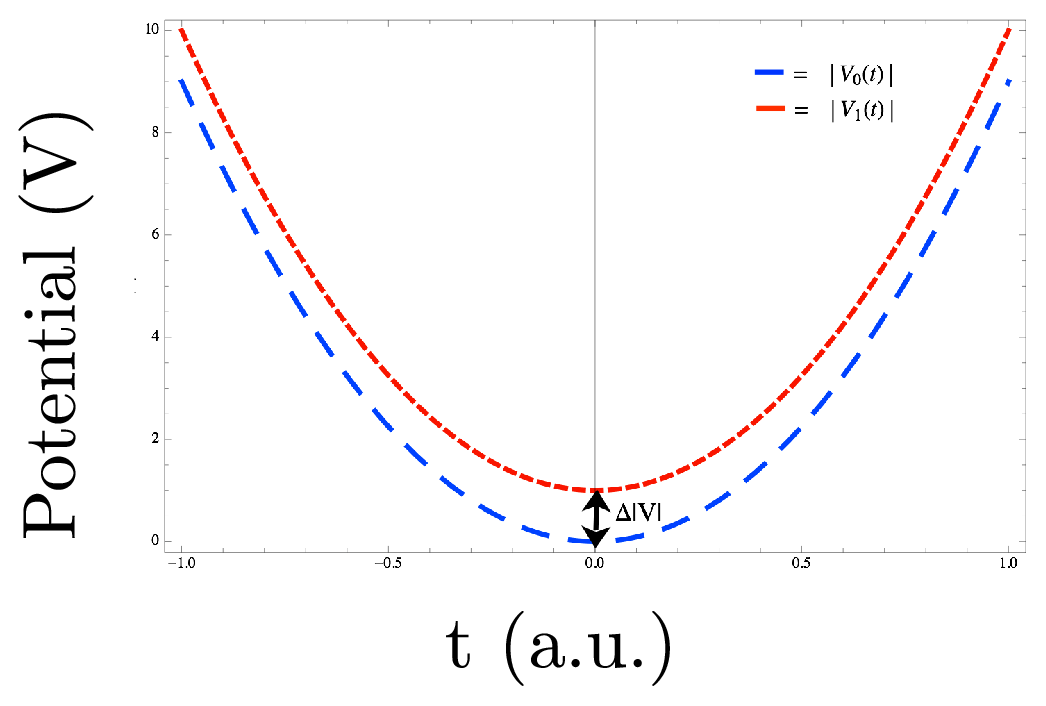, scale=0.50}}} \\ \hline
Quantum gate &Input& Initial potential & Final potential \\ \hline
&&&\\ 
$\textbf{U}_{0}$&$\sigma_{0}$&$V_{0}(t)=i3t\sigma_{0}$&$V_{1}(t)=i3t\sigma_{0}-i\sigma_{z}$\\ 
&&&\\ \hline
\end{tabular}
\fcaption{\label{tab:1w}Resume of $\textbf{U}_{0}$ .}
\end{center}
\vspace{0.15cm}
\end{table}
\end{center}
\indent
\begin{center}
\begin{table}[h]
\begin{center}
\begin{tabular}{|c|c|c|c|}
\multicolumn{4}{c}{ $\textbf{Potential dynamics}$} \\ 
\multicolumn{4}{c}{
\centerline{\epsfig{file=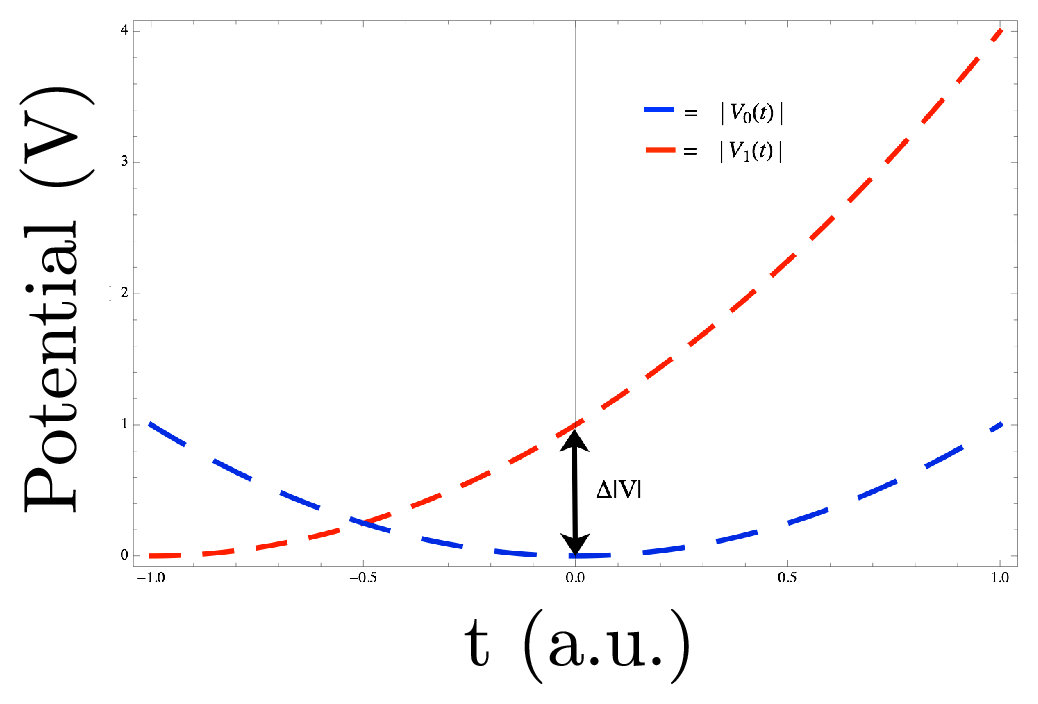, scale=0.50}}} \\ \hline
Quantum gate &Input& Initial potential & Final potential \\ \hline
&&&\\ 
$\textbf{U}_{1}$&$\sigma_{1}$&$V_{0}(t)$&$V_{1}(t)$\\ 
&&$=t(\sigma_{2}-i\sigma_{3})$&$=(t+1)\sigma_{2}-it\sigma_{3}$\\ 
&&&\\ \hline
\end{tabular}
\fcaption{\label{tab:1e}Resume of $\textbf{U}_{1}$ .}
\end{center}
\vspace{0.15cm}
\end{table}
\end{center}
\indent
\begin{center}
\begin{table}[h]
\begin{center}
\begin{tabular}{|c|c|c|c|}
\multicolumn{4}{c}{ $\textbf{Potential dynamics}$} \\ 
\multicolumn{4}{c}{
\centerline{\epsfig{file=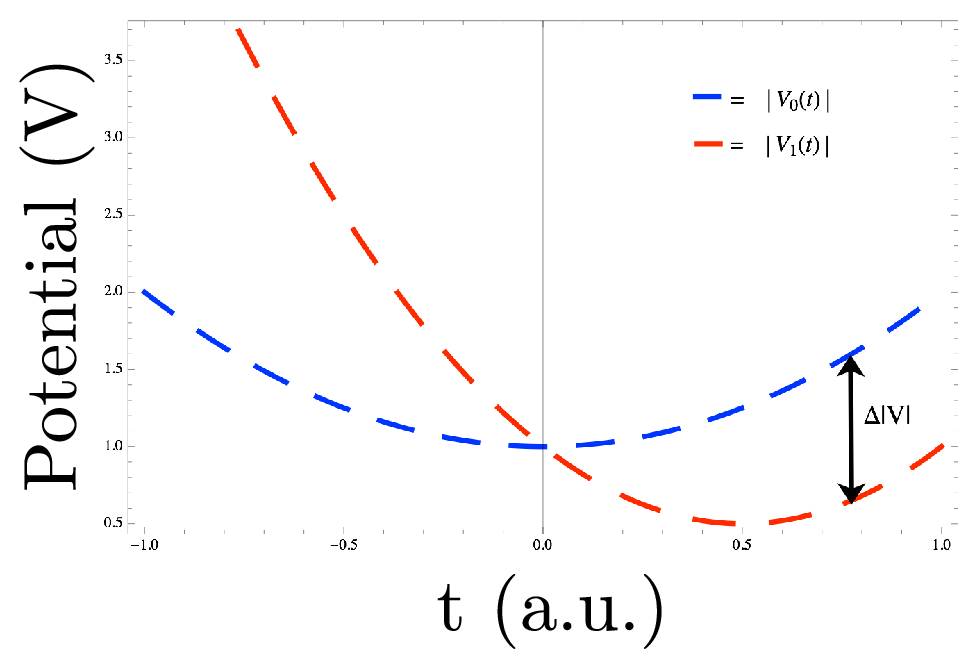, scale=0.50}}} \\ \hline
Quantum gate &Input& Initial potential & Final potential \\ \hline
&&&\\ 
$\textbf{U}_{2}$&$\sigma_{2}$&$V_{0}(t)$&$V_{1}(t)$\\ 
&&$=-t(\sigma_{1}+i\sigma_{3})$&$=-(t+1)\sigma_{1}-it\sigma_{3}$\\ 
&&&\\ \hline
\end{tabular}
\fcaption{\label{tab:1r}Resume of $\textbf{U}_{2}$ .}
\end{center}
\vspace{0.15cm}
\end{table}
\end{center}
\indent
\begin{center}
\begin{table}[h]
\begin{center}
\begin{tabular}{|c|c|c|c|}
\multicolumn{4}{c}{ $\textbf{Potential dynamics}$} \\ 
\multicolumn{4}{c}{
\centerline{\epsfig{file=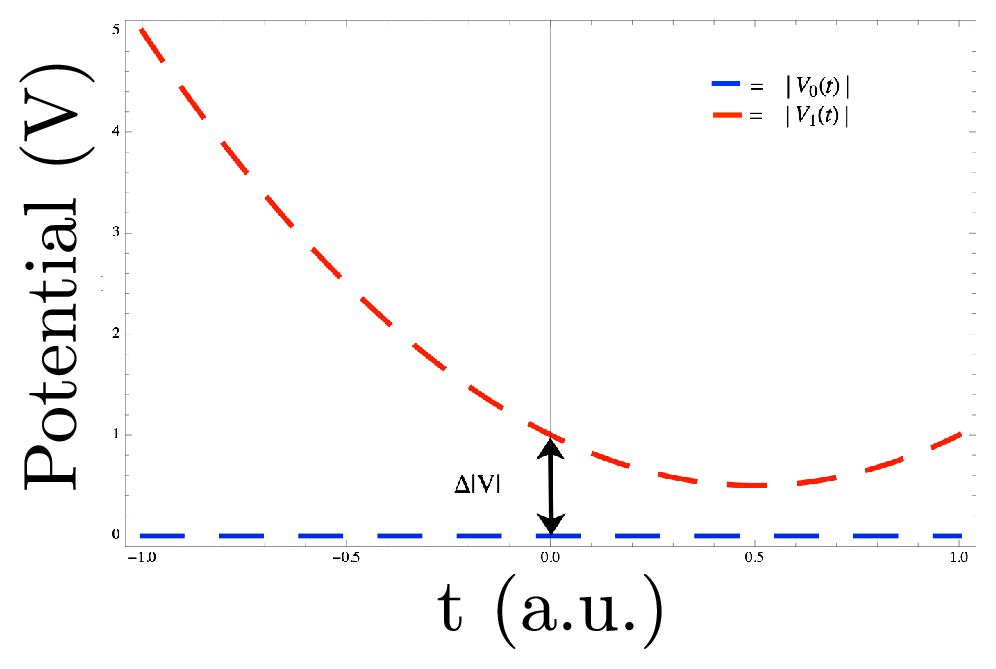, scale=0.50}}} \\ \hline
Quantum gate &Input& Initial potential & Final potential \\ \hline
&&&\\ 
$\textbf{U}_{3}$&$\sigma_{3}$&$V_{0}(t)$&$V_{1}(t)$\\ 
&&$=-it(\sigma_{0}+\sigma_{3})$&$=-i(t+1)\sigma_{0}-it\sigma_{3}$\\ 
&&&\\ \hline
\end{tabular}
\fcaption{\label{tab:1t}Resume of $\textbf{U}_{3}$ .}
\end{center}
\end{table}
\end{center}
\indent
\pagebreak[4]
\subsection[Multiple qubit quantum gates generation by a quantum transistor]{\label{sec:level2a01} Multiple qubit quantum gates generation by a quantum transistor}

Remarkable fact about $\gamma$-matrices is they have similar mathematical structures with Pauli matrices \cite{Poole:1982mo}. Quantum gates for two qubits can also be decomposed into $\gamma$-matrices \cite{PhysRevA.64.024303}.

To demonstrate how the proposed principle works in the case of two qubits, one can use the \textbf{$\gamma$}-matrix\cite{itzykson1985quantum} to describe the complex-four vector as following
\begin{equation}
\label{2qb}
\slashed{a}=\sum\limits_{\mu=0}^{3}a_{\mu}\gamma^{\mu},
\end{equation}
\noindent
where $\slashed{a}$ is a complex-four vector represented in $\gamma$-matrix and symbolized by Feynman slash notation. 

It is also clear that the product of two vectors, $a_{\mu}b^{\mu}$, in the representation of $\gamma$-matrix obeys complex four-vector since the $\gamma$-matrices satisfy Clifford algebra 
\begin{equation}
\label{sati2}
\{\gamma^{\mu},\gamma^{\nu}\}=\gamma^{\mu}\gamma^{\nu}+\gamma^{\nu}\gamma^{\mu}=2\eta^{\mu \nu}.
\end{equation}
\indent
The initial one-dimensional Dirac equation of two relativistic spin qubits \textit{at rest} is represented by
\begin{equation}
\label{2qta}
\bigg(i\gamma_{3}\frac{d}{dt}+V_{0}\bigg)|\{0,1\}^{1}\rangle_{i}=\varepsilon_{0}|\{0,1\}^{1}\rangle_{i},
\end{equation}
\noindent
where $\{V_{0},|\{0,1\}^{1}\rangle_{i}\}$ is a set of initial potential and quantum state of the system consisting two qubits.

The perturbation on the system occurs if the two relativistic spin qubits moving in classical electromagnetic field, therefore the quantum system obeys the new one-dimensional Dirac equation  
\begin{equation}
\label{2qtaf}
\bigg(i\gamma_{3}\frac{d}{dt}+V_{0}+\Delta V\bigg)|\{0,1\}^{2}\rangle_{f}=\varepsilon_{1}|\{0,1\}^{2}\rangle_{f},
\end{equation}
\noindent
where $\{V_{0}+\Delta V,|\{0,1\}^{2}\rangle_{f}\}$ is a set of final potential and quantum state of the system. Similar to in the case of one qubit in Eq. (\ref{rlH}), here the perturbation term corresponds to Lorentz force, therefore $q$ represents the effective charge of the two relativistic spin qubits.

The change of potential and state in the case of two qubits follows the \textit{Theorem 1}, nevertheless the intertwining operator in Eq. (\ref{e0}) is substituted by
\begin{equation}
\label{2qsz}
\hat{\mathcal B}(\{\textbf{C(\textbf{U}$_{i}$)},\textbf{SWAP}\}) = \alpha_{i}(t)\gamma^{0}+(V_{N-1}(t)-\beta_{i}(t))\{\textbf{C(\textbf{U}$_{i}$)},\textbf{SWAP}\},
\end{equation}
\noindent
where \textbf{C(\textbf{U}$_{i}$)} is \textbf{Controlled-U$_{i}$} gate, and the perturbation on Dirac Hamiltonian is defined by
\begin{equation}
\label{nh2}
\Delta V = -i\vec{\gamma_{3}}\wedge\{\vec{\textbf{C(\textbf{U}$_{i}$)}},\vec{\textbf{SWAP}}\}.
\end{equation}

\begin{figure}[h!]
\begin{center}
\centerline{\epsfig{file=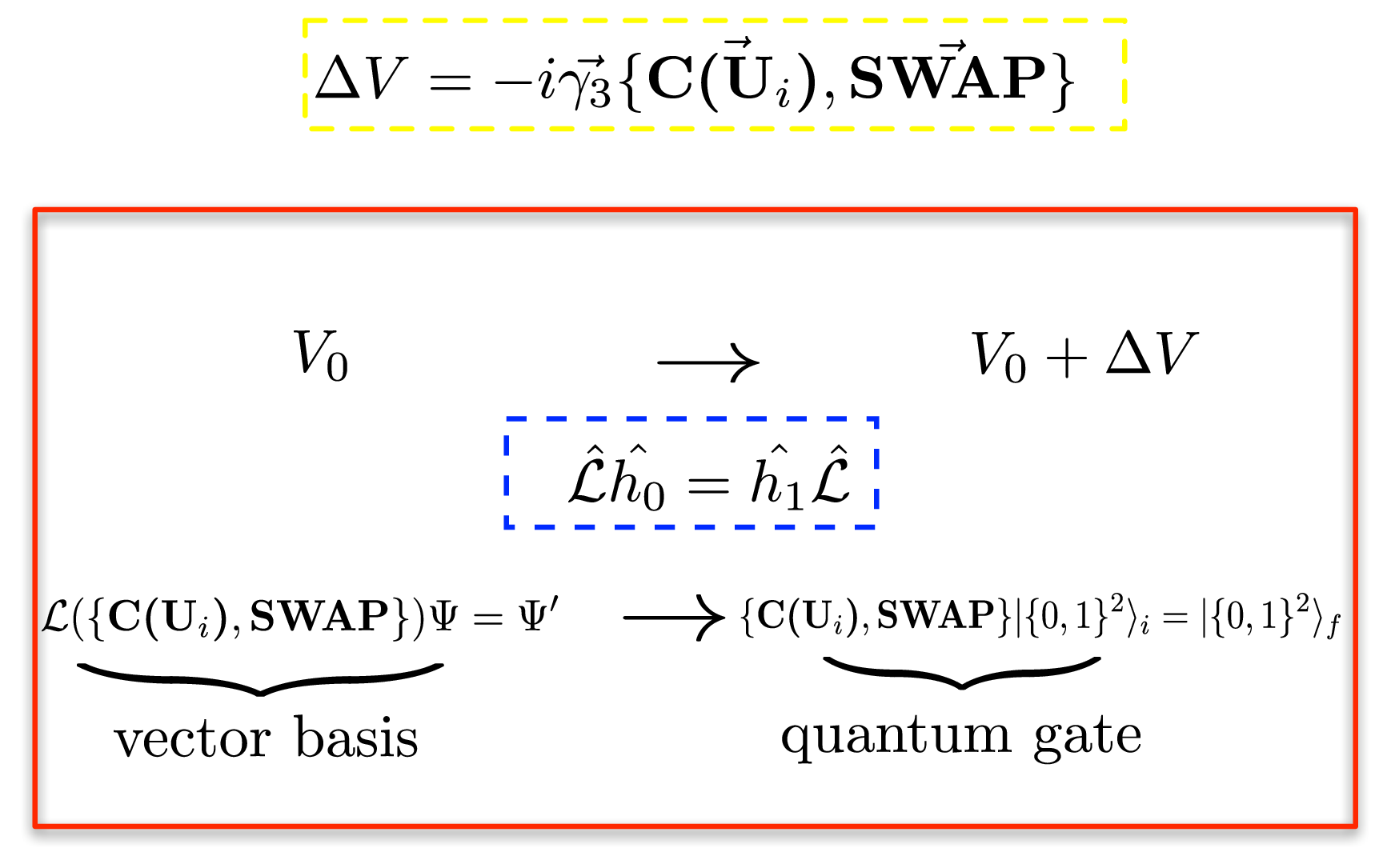, scale=0.40}}
\fcaption{\label{fne} The scheme of methods for two qubits. The calculation inside the red box represents the main method: Darboux transformations on one-dimensional Dirac equation causes the transformation a set of potential and state from the old into the new one, $\{V_{0}, |\{0,1\}^{2}\rangle_{i}\}\rightarrow \{V_{0}+\Delta V, |\{0,1\}^{2}\rangle_{f}\}$. However, if the potential perturbation term is fixed in a certain formulation related to Lorentz force form, as written inside the yellow box, the transformation operator of the quantum state is changed into the form of quantum gates.}
\end{center}
\end{figure}

Similar to the case of a single qubit, the term $\{\vec{\textbf{C(\textbf{U}$_{i}$)}},\vec{\textbf{SWAP}}\}$ in Eq. (\ref{nh2}) corresponds to the direction of an external perturbation, while $\vec{\gamma_{3}}$ belongs to the direction of the two relativistic spin qubits.
This method converts the ``vector basis'' aspect as found in Eq. (\ref{nh2}) into ``quantum gates'' aspect of $\gamma$-matrices, since the initial Dirac potential is vanished by the perturbation term due to the transformation. The scheme of this method is illustrated in Fig. (\ref{fne}).

By similar manner, one can obtain for $n$-qubits, since the vector basis and quantum gates are easily acquired by Pauli-Dirac matrix generator \cite{Poole:1982mo}. The multiple qubits quantum computation on Dirac equation by utilizing Lorentz force can be realized by simply substitution of Pauli matrices in the \textit{Theorem 1} by $\theta_{\mu}$-matrices.

To generalize our proposal, the following set of equations show the multiple qubits quantum computation on Dirac equations utilizing Lorentz force which is transforming the initial set of Dirac equations as given in Eq. (\ref{gde1}) into the final set of new Dirac equations

\begin{figure}[h!]
\centerline{\epsfig{file=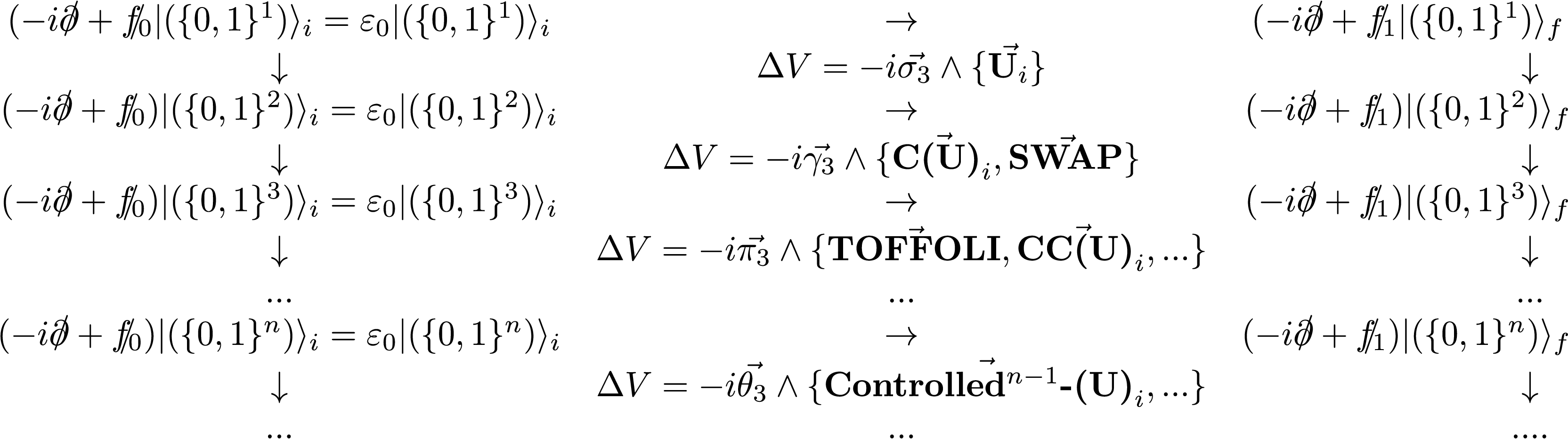, angle=0, scale=0.24}}
\begin{equation}
\label{gde1a}
\end{equation}
\end{figure}


\subsection[Coupling of quantum gates]{\label{sec:level2a01a}Coupling of quantum gates}

According to the previous sections, for instances, the \textbf{Controlled$^{N}$-NOT} can be obtained by the following expressions
\begin{center}
\begin{subequations}
\begin{align}
\textbf{Controlled$^{1}$-NOT} & = \textbf{CNOT}&=\textbf{U}_{0}\bigoplus_{co\{\pm\}}\textbf{U}_{1}  \\
\textbf{Controlled$^{2}$-NOT}&=\textbf{TOFFOLI}&=\textbf{U}_{0}\bigoplus_{co\{\pm\}}\textbf{U}_{0}\bigoplus_{co\{\pm\}}\textbf{U}_{1}\\
\textbf{Controlled$^{N}$-NOT} &&=\overbrace{\textbf{U}_{0}\bigoplus_{co\{\pm\}}...\textbf{U}_{0}}^{N-1}\bigoplus_{co\{\pm\}}\textbf{U}_{1}
\end{align}
\end{subequations}
\end{center}
\indent
Then, for \textbf{NOT-Cyclic$^{N}$}, one can find that
\\
\begin{center}
\begin{subequations}
\begin{align}
\textbf{NOT-Cyclic$^{1}$} &&=\textbf{U}_{1}\bigoplus_{cy\{\pm\}}\textbf{U}_{0}  \\
\textbf{NOT-Cyclic$^{2}$}&&=\textbf{U}_{1}\bigoplus_{cy\{\pm\}}\textbf{U}_{0}\bigoplus_{cy\{\pm\}}\textbf{U}_{0}\\
\textbf{NOT-Cyclic$^{N}$} &&=\textbf{U}_{1}\bigoplus_{cy\{\pm\}}\overbrace{\textbf{U}_{0}\bigoplus_{cy\{\pm\}}...\textbf{U}_{0}}^{N-1}
\end{align}
\end{subequations}
\end{center}
\indent

\begin{figure*}[p!]
\begin{center}
\subfigure [~Illustration for $\textbf{Controlled-NOT}$.]{
\centerline{\epsfig{file=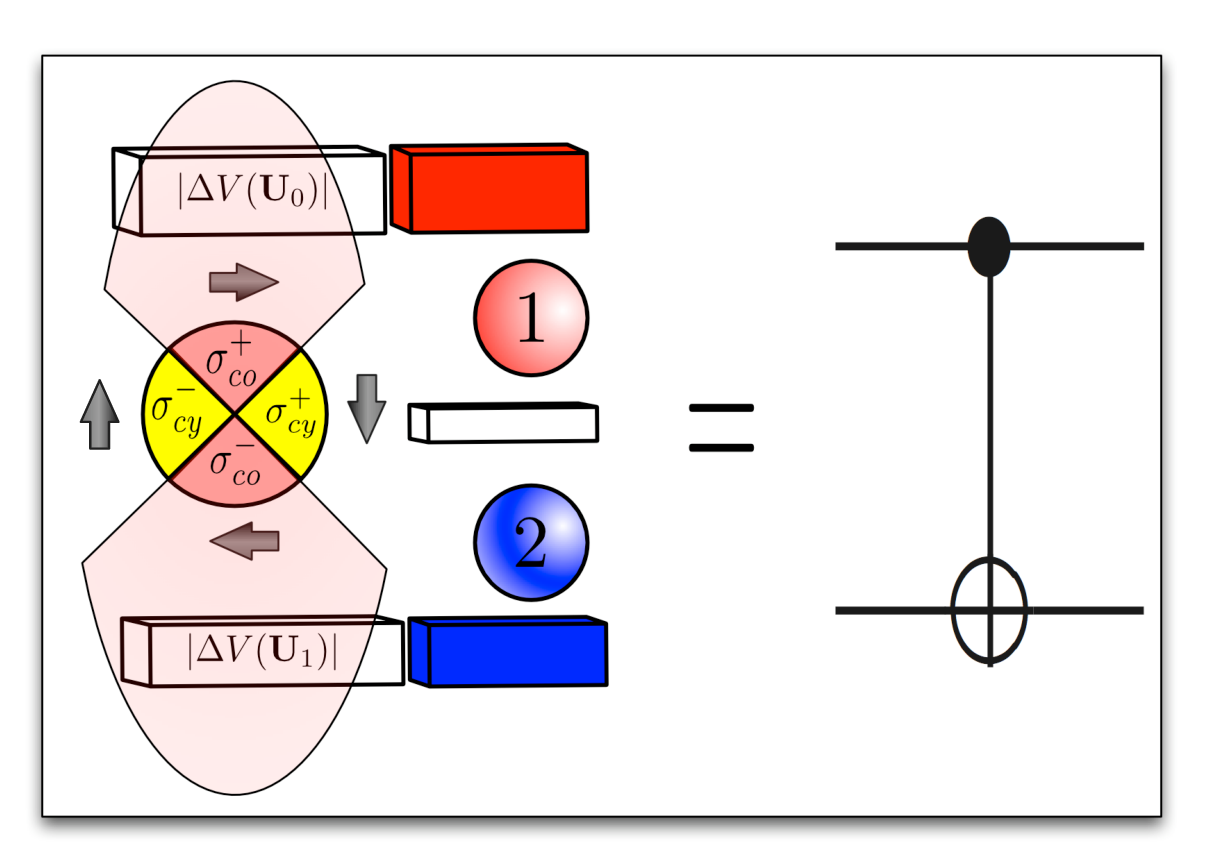, scale=0.40}}

\label{sub1a}
}
\subfigure[~Illustration for $\textbf{NOT-Cyclic}$.]{
\label{fig:sub:1b}
\centerline{\epsfig{file=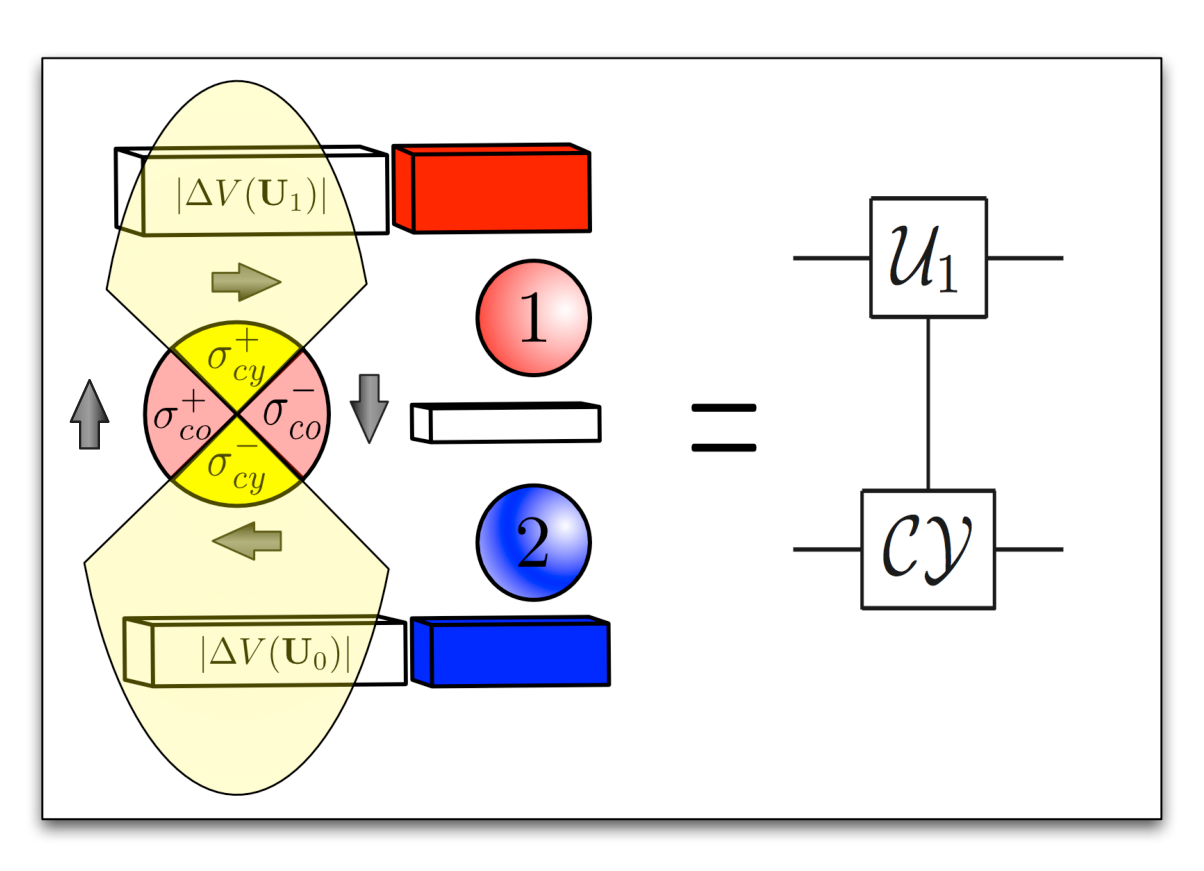, scale=0.40}}

}
\fcaption{\label{fig:qm1} The illustration of our mathematical proposal for describing the process of quantum computation due to the perturbation of the target qubit by an external force. The target qubit performs $\text{Unitary}$ gate due to the perturbation. One can realize $\textbf{Controlled-NOT}$ by setting the control operator, while for $\textbf{NOT-Cyclic}$ by cyclic operator on the coupling.}
\end{center}
\end{figure*}

\subsection[Physical interpretation]{\label{sec:level2tu}Physical interpretation of control-, cyclic-, and target-operators}

In this section, we present a new interpretation of the control-, cyclic-, and target-operators in quantum gates based upon the explanation in the previous sections. The illustrations are given in Fig. (\ref{fig:qm1}) for the quantum gates of two qubits.

Under this scheme, target operator acts on each qubit and is coupled by $\{\text{control, cyclic}\}$-operators for constructing $n$-qubits quantum gates. 

It is shown in Fig. (\ref{sub1a}) that when the control operator is turned on, the circuit is in \textbf{Controlled-NOT} form and it is mathematically expressed by Eq. (\ref{eq0a}). 

As can be seen in Fig. (\ref{fig:sub:1b}) that the coupling operator is changeable, then, the turning cyclic operator on causes the gate in \textbf{NOT-Cyclic} form where it is represented by Eq. (\ref{q1}).  

Another interesting feature of this scheme is that, for a single qubit, both positive and negative energy, $\{\psi^{(+)},\psi^{(-)}\}$, can be used one-qubit quantum computation. As has been discussed in Sec. (\ref{sec:level1}), the degeneracy is broken for the case of multiple qubits: the positive (negative) energy, $\psi^{(+)}$ ($\psi^{(-)}$), is the signature of $|0\rangle$ ($|1\rangle$) state of the first qubit. Therefore, swapping the quantum state of multiple qubits is simultaneously interchange the energy of the system under the action of $\{\textbf{SWAP, Full SWAP}\}$ gate which is realized by the setting of $cyclic$-operator on the coupling, while the chose of \textit{control}-operator ensures the energy conservation.

\section[Conclusion]{\label{sec:level5}Conclusion}

In this paper, we propose a modification form of a Dirac equation by allowing the potential in complex four-vector form representing the classical perturbation on a quantum system. The notion of this work is similar as proposed in Ref. \cite{zakharov1968stability}: the modified Schr\"{o}dinger equation form can be used to represent nonlinear phenomena of a classical system such as water wave, by allowing the potential and also its perturbation are in the classical form.

We have shown that $n$-relativistic spin qubit in external classical electromagnetic fields can be used for resource of quantum transistor. The mathematical explanation behind the proposal is on the action of Darboux transformation on one-dimensional Dirac equation at which the potential is a complex Dirac four-potential utilizing the Lorentz force as perturbing potential. These mathematical methods take advantage of the \textit{vector-quantum gates duality} of Pauli matrices.  

It is also shown that our proposal can closely describe the resources of \textit{quantum} transistor required for implementing quantum computation: every \textit{n}-qubits require \textit{n}-quantum transistors. In this scheme, the transistor is prepared with certain initial potential function, $V_{initial}$, then the final potential function, $V_{final}$, is obtained to generate certain quantum gate. 

However, the proposed scheme as presented is aimed as a primitive approach of a quantum transistor based upon Dirac equations. The next challenge is how to implement it into a more specific and complex physical system. The correlation between a specific physical quantity, as though quantum Rabi oscillations in cavity quantum electrodynamics and bandgap in various types of graphene transistor, and the change of qubit state due to the tuning of the direction of the external electromagnetic field, shall be explored.

The proposal may be also very useful for explaining the quantum system used not only the magnitude but also direction of a physical quantity to encode the qubits, as found in superconducting flux qubits, where the two distinguishable configurations are obtained by induced clockwise or anti clockwise current \cite{IChio2003}. 

The proposed formalism opens the possibility to involve the relativity into the further study of quantum computation and information, coincidentally with an emerging assertion that closed timelike curve can solve classical problems in computation such as \textbf{NP}-complete problems \cite{PhysRevD.44.3197,PhysRevA.70.032309,brun2003computers,aaronson2009closed}.

For the next work, it is also very interesting to consider if quantum circuit is constructed under the Dirac equations for $n$-qubits instead of conventional way in Schr\"{o}dinger representations. The similar evaluation as provided in Ref. \cite{trisetyarso2009circuit} may give noteworthy results since this kind of construction may influence the depth and space resource of a quantum circuit. In this work, we assume that intertwining operation runs only the potential of relativistic spin qubits. The computation of a qubit still runs under linear transformation. If the intertwining operation runs both on the transformation of potential and qubit, this may contribute into the theory of the topological quantum computing\cite{kauffman2004braiding} on a massless Dirac-fermion.
\\
\indent
\\
$\textit{Acknowledgments}$ - This work was supported in part by Grant-in-Aid for Scientific Research by MEXT, Specially Promoted Research No. 18001002 and in part by Special Coordination Funds for Promoting Science and Technology. We also would like to thank Prof. Kohei M. Itoh, Rodney Van Meter Ph.D, and Munawar Riyadi (UTM, Malaysia) for fruitful discussion. 

\nonumsection{References}

\end{document}